\documentclass{ieeetmlcn}

\usepackage{amsmath,amsfonts}
\usepackage{algorithmic}
\usepackage{algorithm}
\usepackage{array}
\usepackage[caption=false,font=normalsize,labelfont=sf,textfont=sf]{subfig}
\usepackage{textcomp}
\usepackage{stfloats}
\usepackage{url}
\usepackage{verbatim}
\usepackage{graphicx}
\graphicspath{{./figs/}}
\usepackage{cite}
\usepackage{hyperref}
\hypersetup{hidelinks=true}
\usepackage{tikz}
\usepackage{soul}
\usetikzlibrary{
	calc,
	shapes,
	fit,
	positioning,
	matrix,
	trees,
	decorations.markings,
	decorations.pathreplacing,
	chains,
}
\usepackage{pgfplots}
\pgfplotsset{compat=1.18}
\usepackage[acronyms,toc]{glossaries}
\usepackage[mode=buildmissing]{standalone}
\usepackage{xcolor}
\usepackage{bm}
\usepackage{makecell}

\definecolor{gray1}{RGB}{237,242,245}
\definecolor{gray2}{RGB}{190,200,210}
\definecolor{gray3}{RGB}{152,162,174}
\definecolor{gray4}{RGB}{77,87,102}
\definecolor{gray5}{RGB}{39,49,66}

\newcolumntype{P}[1]{>{\raggedright\arraybackslash}p{#1}}
\def\authorrefmark#1{\ensuremath{^{\textbf{#1}}}}

\def\BibTeX{{\rm B\kern-.05em{\sc i\kern-.025em b}\kern-.08em
    T\kern-.1667em\lower.7ex\hbox{E}\kern-.125emX}}
\AtBeginDocument{\definecolor{tmlcncolor}{cmyk}{0.93,0.59,0.15,0.02}\definecolor{NavyBlue}{RGB}{0,86,125}}
\def\OJlogo{\vspace{-4pt}\includegraphics[height=18pt]{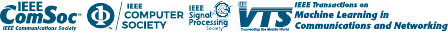}}
\def\seclogo{\vspace{10pt}\includegraphics[height=18pt]{logo/new_logo.pdf}}
\def\authorrefmark#1{\ensuremath{^{\textbf{#1}}}}

\begin{document}

\receiveddate{XX Month, XXXX}
\reviseddate{XX Month, XXXX}
\accepteddate{XX Month, XXXX}
\publisheddate{XX Month, XXXX}
\currentdate{XX Month, XXXX}
\doiinfo{TMLCN.2022.1234567}

\markboth{}{A.~Pastore, A.~Agustín, Á.~Valcarce}

\title{Optimizing Wireless Discontinuous Reception via MAC Signaling Learning}

\author{
	Adriano Pastore\authorrefmark{1}, Senior Member, IEEE,
	Adrián Agustín de Dios\authorrefmark{1}, Senior Member, IEEE,
	Álvaro Valcarce\authorrefmark{2}, Senior Member, IEEE
}
\affil{Centre Tecnològic de Telecomunicacions de Catalunya (CTTC), Av.~Carl Friedrich Gauss 7, 08860 Castelldefels, Spain}
\affil{Nokia Networks France, 12 Rue Jean Bart, 91300 Massy, France}
\corresp{Corresponding author: Adriano Pastore (email: adriano.pastore@ cttc.cat).}
\authornote{This work has been funded by Nokia through its program of academic partnerships.}


\newacronym{3GPP}{3GPP}{3rd Generation Partnership Project}
\newacronym{5G}{5G}{Fifth Generation}
\newacronym{5G NR}{5G NR}{5G New Radio}
\newacronym{5GQI}{5GQI}{5G QoS Indicator}
\newacronym{ACK}{ACK}{Acknowledgement}
\newacronym{AI}{AI}{Artificial Intelligence}
\newacronym{BLER}{BLER}{Block Error Rate}
\newacronym{BTS}{BTS}{Base Transceiver Station}
\newacronym{C-RNTI}{C-RNTI}{Cell Radio Network Temporary Identifier}
\newacronym{CE}{CE}{Control Element}
\newacronym{CSI}{CSI}{Channel State Information}
\newacronym{CQI}{CQI}{Channel Quality Indicator}
\newacronym{DQN}{DQN}{Deep Q-Network}
\newacronym{DCI}{DCI}{Downlink Control Information}
\newacronym{DL}{DL}{Downlink}
\newacronym{DL-SCH}{DL-SCH}{Downlink Shared Channel}
\newacronym{DRX}{DRX}{Discontinuous Reception}
\newacronym{ERM}{ERM}{Experience Replay Memory}
\newacronym{gNB}{gNB}{gNodeB}
\newacronym{HARQ}{HARQ}{Hybrid ARQ}
\newacronym{KPI}{KPI}{Key Performance Indicator}
\newacronym{L2C}{L2C}{Learning to Communicate}
\newacronym{LCID}{LCID}{Logical Channel ID}
\newacronym{LTE}{LTE}{Long Term Evolution}
\newacronym{MAC}{MAC}{Medium Access Layer}
\newacronym{MARL}{MARL}{Multi-Agent Reinforcement Learning}
\newacronym{MCS}{MCS}{Modulation and Coding Scheme}
\newacronym{ML}{ML}{Machine Learning}
\newacronym{MNO}{MNO}{Mobile Network Operator}
\newacronym{PDB}{PDB}{Packet Delay Budget}
\newacronym{PDCCH}{PDCCH}{Physical Downlink Control Channel}
\newacronym{PDCP}{PDCP}{Packet Data Convergence Protocol}
\newacronym{PDSCH}{PDSCH}{Physical Downlink Shared Channel}
\newacronym{PDU}{PDU}{Protocol Data Unit}
\newacronym{PHY}{PHY}{Physical Layer}
\newacronym{PMI}{PMI}{Precoder Matrix Indicator}
\newacronym{PUCCH}{PUCCH}{Physical Uplink Control Channel}
\newacronym{PUSCH}{PUSCH}{Physical Uplink Shared Channel}
\newacronym{QoS}{QoS}{Quality of Service}
\newacronym{RACH}{RACH}{Random Access Channel}
\newacronym{RAR}{RAR}{Random Access Response}
\newacronym{RI}{RI}{Rank Indicator}
\newacronym{RL}{RL}{Reinforcement Learning}
\newacronym{RLC}{RLC}{Radio Link Control}
\newacronym{RRC}{RRC}{Radio Resource Control}
\newacronym{RSRP}{RSRP}{Reference Signal Received Power}
\newacronym{SAW}{SAW}{Stop-And-Wait}
\newacronym{SSSG}{SSSG}{Search Space Set Group}
\newacronym{SDU}{SDU}{Service Data Unit}
\newacronym{SINR}{SINR}{Signal-to-Interference-and-Noise Ratio}
\newacronym{SNR}{SNR}{Signal-to-noise Ratio}
\newacronym{SS}{SS}{Synchronization Signal}
\newacronym{TB}{TB}{Transport Block}
\newacronym{TBS}{TBS}{Transport Block Size}
\newacronym{TDD}{TDD}{Time Division Duplexing}
\newacronym{TS}{TS}{Technical Specification}
\newacronym{TTI}{TTI}{Transmission Time Interval}
\newacronym{UCI}{UCI}{Uplink Control Information}
\newacronym{UE}{UE}{User Equipment}
\newacronym{UL}{UL}{Uplink}
\newacronym{UL-SCH}{UL-SCH}{Uplink Shared Channel}
\newacronym{XR}{XR}{Extended Reality}
\newacronym{VR}{VR}{Virtual Reality}
\newacronym{FTP}{FTP}{File Transfer Protocol}

\begin{abstract}
We present a \gls{RL} approach to the problem of controlling the \gls{DRX} policy from a \gls{BTS} in a cellular network.
We do so by means of optimally timing the transmission of fast Layer-2 signaling messages (a.k.a.~\gls{MAC} \glspl{CE} as specified in 5G New Radio).
Unlike more conventional approaches to \gls{DRX} optimization, which rely on fine-tuning the values of \gls{DRX} timers, we assess the gains that can be obtained solely by means of this MAC CE signalling.
For the simulation part, we concentrate on traffic types typically encountered in Extended Reality (XR) applications, where the need for battery drain minimization and overheating mitigation are particularly pressing. Both 3GPP \gls{5G NR} compliant and non-compliant (``beyond 5G'') \gls{MAC} \glspl{CE} are considered. Our simulation results show that our proposed technique strikes an improved trade-off between latency and energy savings as compared to conventional timer-based approaches that are characteristic of most current implementations. Specifically, our RL-based policy can nearly halve the active time for a single \gls{UE} with respect to a na\"{i}ve \gls{MAC} \gls{CE} transmission policy, and still achieve near 20\% active time reduction for 9 simultaneously served \glspl{UE}.

\end{abstract}


\maketitle

\section{INTRODUCTION}
\IEEEPARstart{T}{he} \gls{MAC} layer protocol specifications up to \gls{5G} networks, as well as all earlier predecessor standards such as \gls{LTE}, have been engineered by hand, mostly following expert guidelines and heuristics. As a promising alternative, recent advances in \gls{MARL} have opened up new possibilities for an experience-driven approach to protocol design \cite{Valcarce21}.

Specifically, in this work, we apply a \gls{MARL} approach to the design of the \gls{DRX} control policy.
\gls{DRX} is a network function that lets \glspl{UE} switch intermittently between an active and an inactive (a.k.a.~sleep) state.
During the inactive state, \glspl{UE} cease to monitor and decode the \gls{PDCCH}, thus saving some energy in the process.
However, this comes at the cost of increased end-to-end latency and a lower \gls{CSI} quality due to missed opportunities for channel reporting.

\gls{DRX} is an important technique for reducing battery drain in mobile devices, as well as for improving the energy efficiency of \gls{5G} radio links. In fact, power-saving efforts in \gls{5G} communications (and beyond) are crucial, since energy efficiency is a requirement that transversally impacts all use cases (eMBB, mMTC, URLLC) \cite{Dahlman24}.
As highlighted in~\cite{Ruy2020}, \gls{PDCCH} monitoring is a primary contributor to UE power consumption~\cite{zte18}. In fact, for common traffic types, the overhead due to \gls{PDCCH} activity is disproportionately large compared to the energy expended on payload data via the \gls{PDSCH}.
For instance, according to \cite[Fig.~8]{Ruy2020}, under an FTP-3 traffic model, the time during which the \gls{UE} \emph{solely} processes the \gls{PDCCH} accounts for 92\% of the total power consumption, while time spent processing the \gls{PDSCH} only contributes 6\% (with the remaining 2\% representing sleep).
This asymmetry is largely due to the \emph{blind decoding} process that \glspl{UE} perform to obtain the \gls{PDCCH}.
In addition, unlike at the \gls{BTS}, energy inefficiencies have multifaceted impacts on the \gls{UE} performance and the comfort of the human user.
This is mostly due to the prevalence of battery-powered devices, heating effects, etc.
Optimizing \gls{DRX} performance is, therefore, a key priority for \glspl{MNO} to ensure that the \glspl{UE} have enough battery to get through the day.

Most attempts at \gls{DRX} optimization have focused on finding the best values for the set of timers that steer the \gls{5G} \gls{DRX} function.
These are namely the \emph{drx-InactivityTimer}, the \emph{drx-onDurationTimer}, and the \emph{drx-onDuration} parameter (see \cite{5GMAC} for details on these).
More recently, signaling-based approaches, such as \emph{\gls{SSSG} switching} and \emph{PDCCH skipping} were standardized in \gls{3GPP} Rel-17, as described in \cite{9771950}. However, usage of these is yet to reach live networks, mainly due to difficulties in finding a logic to steer them. Fortunately, the abundance of data that can be obtained in cellular networks paves the way for \gls{ML} approaches to this problem.
Authors in \cite{Muda19} and \cite{Zhou2019}, for example, train traffic predictors and then derive the timer values from the obtained forecasts.
On the other hand, \cite{Ericsson21} proposes a contextual bandit and non-standard 5G signaling to select among a set of predefined \gls{DRX} configurations.

Given the inherent difficulties in producing truly general traffic forecasts, and our practical wish to remain within \gls{5G}-compliant signaling, we set off to optimizing \gls{DRX} exclusively using the tools available in today's cellular networks.
A potential approach to this could be to treat the optimization of \gls{DRX} timers as a black-box optimization problem. However, \gls{5G} networks only permit updating the value of these timers via Layer 3 signaling, which requires slow and heavy RRC reconfigurations. This is not flexible or fast enough to leverage the energy-saving opportunities that modern lightweight devices require.
For these reasons, we decided to exploit some underused L2 signaling opportunities.
To this end, we devised a \gls{RL} approach compatible with \gls{5G NR} specifications such as those in recent releases of \gls{3GPP} \gls{TS} 38.321, and compare it to the state-of-the-art baselines specified in those standards.
To the best of our knowledge, this is the first reported usage of \gls{ML} to enhance \gls{UE} power savings via low-layer signaling.

Specifically, these baselines are rule-based policies that rely on a timers-based logic.
This logic, described in Section~\ref{ssec:timers}, controls a collection of timers that will count down and be reset depending on traffic demands.
The timer logic is replicated at both ends of the \gls{BTS}-\glspl{UE} links, to guarantee \gls{BTS} and \glspl{UE} timers synchronization.

Alternatively to this timers-based logic, the \emph{start} of an inactive phase may also be triggered by the \gls{BTS} via the transmission of a dedicated \gls{MAC} \gls{CE}. The \emph{termination} of inactive phases, in turn, is still dictated by time-out events.
This intervention of the \gls{BTS} via \gls{CE} transmissions prompts an inactivity timer to be reset and thus partly overrides the ``fallback'' logic of timers. These \glspl{CE} are standardized, but their usage is not widespread in commercial implementations.
Our simulations suggest that \gls{BTS}-side \gls{RL} policies for timing the transmission of \glspl{CE} can be used effectively to improve the \gls{DRX}-induced power savings under stringent latency requirements.

\section{SYSTEM MODEL}   \label{ssec:DRX_single_user}
\begin{figure}[ht!]
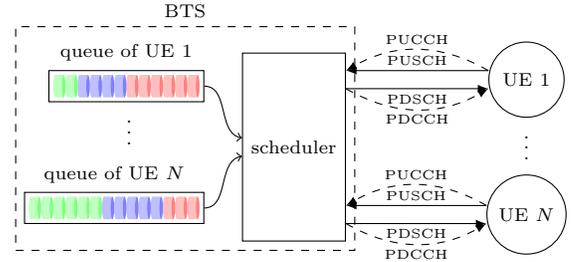

	\centering
	\includestandalone{tikz/two-user_system}
	\caption{Wireless network architecture.}
	\label{fig:DRX_multi_user}
\end{figure}

We consider a downlink setting involving one \gls{BTS} and one or more \glspl{UE}, as sketched in Figure~\ref{fig:DRX_multi_user}.

The \gls{UE} queues store a collection of \gls{RLC} \glspl{PDU} that are assembled into \gls{MAC} \glspl{SDU}, or so-called \glspl{TB}. We assume that one \gls{TB} at most can be transmitted by the \gls{BTS} in each \gls{TTI}, and that the \gls{TTI} duration remains fixed to $1$ millisecond throughout. 
The \glspl{UE} can communicate with the \gls{BTS} through an error-free control channel, \gls{PUCCH}, allowing to acknowledge or not the received \gls{TB}.
The scheduler is responsible for assigning the radio resources to the \glspl{UE} taking into account which \glspl{UE} have information in their respective downlink queues and which ones are active, listening to the \gls{PDCCH}.
We assume that the scheduler is given, for example, based on Round Robin, and we cannot modify its policy.
Every time the scheduler takes a decision, it informs the respective \gls{UE} through the \gls{PDCCH}.

At each \gls{UE}, we distinguish the following states:
\begin{itemize}   \itemsep0em
    \item   If $W_u(t) = 1$ (active), the $u$-th \gls{UE} tries to decode the \gls{PDCCH} in \gls{TTI} $t$ to elucidate if a \gls{TB} is transmitted in the \gls{PDSCH}.
    \item   If $W_u(t) = 0$ (inactive), the $u$-th \gls{UE} does not try to decode the \gls{PDCCH} in \gls{TTI} $t$. This is the equivalent of a sleep or energy-savings state at the \gls{UE}.
\end{itemize}

\subsection{DRX policy based on timers}   \label{ssec:timers}

\begin{figure}[ht]
	\centering
	\includestandalone[scale=0.6]{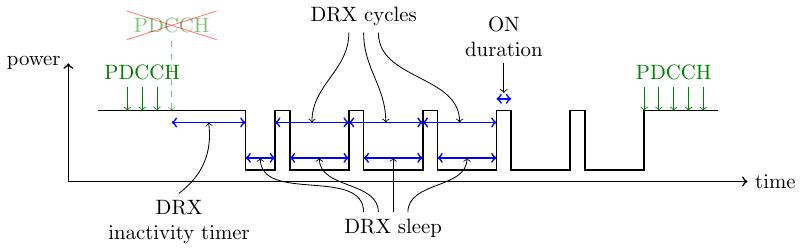}
	\caption{Power consumption profile over time for a \gls{UE} operated with timer-based \gls{DRX}.}
	\label{fig:DRX_power_consumption_over_time}
\end{figure}

In \gls{3GPP} \gls{5G NR}, inactive period durations are determined by timers, providing the \gls{BTS} with precise knowledge of each UE's status for scheduling considerations.
3GPP defines the following parameters to configure the timers:
\begin{itemize}
	\item	\textbf{\texttt{DRXCycle}}. The total duration during which a \gls{UE}'s DRX timer is valid (active and inactive periods).
	\item	\textbf{\texttt{onDurationTimer}}. The part of the \emph{DRXCycle} during which the \gls{UE} is active.
	\item	\textbf{\texttt{drx-onInactivityTimer}}. This timer keeps the \gls{UE} active after receiving a message. Upon message reception, the timer is reset. Once it expires, the \gls{UE} transitions to an inactive state until the current \gls{DRX} cycle concludes.
\end{itemize}
The general working principle, with some simplifications, is best illustrated by Figure~\ref{fig:DRX_power_consumption_over_time}.
We assume that the \gls{UE} is actively monitoring the \gls{PDCCH} and receiving downlink data.
Then, at some point, this \gls{UE}'s downlink buffer is depleted, and the scheduler ceases to notify the \gls{UE} about new data.
The \gls{UE} continues to monitor the \gls{PDCCH} during a number of consecutive \glspl{TTI} denoted by \textbf{\texttt{drx-onInactivityTime}}.
When the inactivity timer expires (i.e., it counts down to zero), the \gls{UE} switches to a periodic behavior that alternates between short listening periods (``ON duration'' defined by \textbf{\texttt{onDurationTimer}}) and prolonged inactivity periods (``\gls{DRX} sleep'' in Figure~\ref{fig:DRX_power_consumption_over_time}) that end at each start of a new DRX cycle.

\subsection{Traffic model and L2 queue management}   \label{ssec:packet_arrivals}

\begin{figure}[ht]
	\centering
	\includestandalone[scale=0.55]{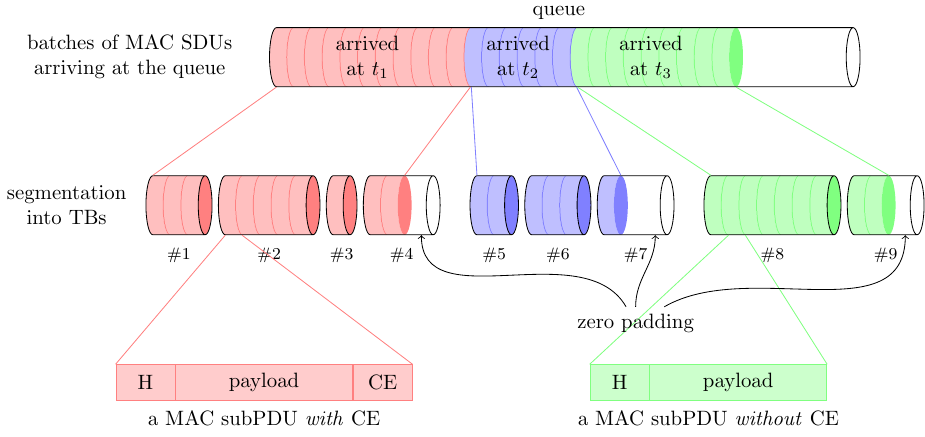}
	\caption{Illustration of the \gls{TB} generation. In this example, the queue contents are collections of \gls{MAC} \glspl{SDU} that have arrived at \glspl{TTI} $t_1$, $t_2$ and $t_3$, respectively. They get segmented by the \gls{MAC} layer into a sequence of \glspl{TB} of different sizes (depending on the channel conditions). Zero padding is applied to fill out the last \gls{TB} corresponding to each batch of payload data.}
	\label{fig:SDU_segmentation}
\end{figure}

At the \gls{BTS}, a stochastic process governs the arrival of user data to the \gls{MAC} as a sequence of \gls{RLC} \glspl{PDU} (cf.\ Figure~\ref{fig:SDU_segmentation}).
In simulations, we have modeled this arrival process to mimic \gls{XR} traffic, which is representative of augmented reality and cloud gaming use cases.
Such type of traffic is well known to have extreme throughput and latency requirements. Furthermore, reducing the power consumption of wearable devices such as \gls{VR} headsets is essential to mitigate user discomfort for overheating reasons.
Minimizing battery drain is also a highly relevant target~\cite{Akyi22,PetGapParMarPed22}.

\begin{table}[ht]
\caption{XR traffic model as described in~\cite{PetGapParMarPed22}.}
\label{table:XR_traffic}
\begin{center}
\footnotesize
\renewcommand\cellalign{lc}
\begin{tabular}{|l|l|}
\hline
Interarrival time: & \makecell{ 16.6 ms \\ (corresponds to 60 frames per second) } \\
\hline
Average data rate: & 60 Mbit/s \\
\hline
\makecell{Average\\packet size:} & \makecell{ 1 Mbit \\ (equals data rate/fps) } \\
\hline
Packet size: & \makecell{ truncated Gaussian with: \\ \hspace{5mm} standard dev.\ = 10.5\% of the mean \\ \hspace{5mm} Min = 50\% of the mean \\ \hspace{5mm} Max = 150\% of the mean } \\
\hline
Jitter: & \makecell{ truncated Gaussian with \\ \hspace{5mm} standard dev.\ = 2 ms \\ \hspace{5mm} Min = 4 ms \\ \hspace{5mm} Max = 4 ms } \\
\hline
\end{tabular}
\end{center}
\end{table}

For these reasons, \gls{XR} traffic is an ideal traffic type to study the performance of \gls{DRX} solutions. A thoroughly validated mathematical model of \gls{XR} traffic has been proposed in~\cite{PetGapParMarPed22} and is summarized in Table~\ref{table:XR_traffic}.
In contrast to other well-known models like FTP3 traffic, the \gls{XR} traffic pattern is of quasi-periodic nature (interarrival times have low variance, and stay stable up to jitter effects), which gives this traffic type a certain level of predictability and as a consequence, substantial potential for \gls{DRX}-managed power savings.

We follow a standard \gls{5G NR} stacked architecture, where downlink \gls{RLC} \glspl{PDU} are handled as \glspl{SDU} by the \gls{MAC} layer.
These are accumulated in a separate queue for each user.
The \gls{MAC} layer may then assemble multiple single-\gls{UE} \glspl{SDU} into a single \glspl{TB} (with possible insertion of \gls{MAC} \glspl{CE}).
The \gls{TB} size is selected to be commensurate with the channel conditions informed by the \gls{PHY} layer (cf.\ Subsection~\ref{ssec:phy_modeling}).\footnote{For simplicity, we will assume that the size of batches of \gls{MAC} \glspl{SDU} entering the queue, as well as the size of \glspl{TB}, are arbitrary integer-valued (in bits).}
The \glspl{TB} are then conveyed from the \gls{BTS} \gls{PHY} to the \gls{UE} via the \gls{PDSCH}, and \gls{MAC} \glspl{SDU} are deleted from the queue only after the \gls{BTS} receives a corresponding \gls{ACK} message from the \gls{UE} via the \gls{PUCCH}.

The \glspl{CE} are reserved fields that can be appended to any single \gls{MAC} sub\gls{PDU} (cf.\ Figure~\ref{fig:SDU_segmentation}). We distinguish between legacy \glspl{CE} (that are part of the \gls{5G NR} technical specifications) and new, additional \glspl{CE} whose semantics we may customize to our needs.
The 3GPP standard defines two legacy \glspl{CE}:
\begin{itemize}
	\item	\textbf{\texttt{DRX\_command\_MAC\_CE}} (\gls{LCID} 60) instructs the \gls{UE} to finish the current active time and enter into a short DRX cycle.
	\item	\textbf{\texttt{Long\_DRX\_command\_MAC\_CE}} (\gls{LCID} 59) instructs the \gls{UE} to finish the current active time and start a long DRX cycle.
\end{itemize}
For simplicity, we consider only the \gls{CE} with \gls{LCID} 59, i.e., we restrict ourselves to long DRX cycles.
Note that these \glspl{CE} partly override the timer logic, in the sense that they command the \gls{UE} to enter inactive mode immediately, instead of waiting for a timeout of the DRX inactivity timer, as in conventional timers-based \gls{DRX}.

\subsection{Performance indicators}   \label{ssec:KPIs}

We will focus on two \glspl{KPI}, namely \emph{activity} and \emph{latency}.
The former is a proxy to the power consumption of a \gls{UE} and we define it as the average fraction of time that the $u$-th \gls{UE} has been active during a window of $T$ \glspl{TTI}:
\begin{equation}
    \langle W_u \rangle
    = \frac{1}{T} \sum_{t=1}^T W_u(t).
\end{equation}
The latter is evaluated as an average \gls{SDU} delay
 as follows:
 Let $D_u^{(i)}$ denote the delay of the $i$-th downlink \gls{SDU} for user $u$, that is, the number of \glspl{TTI} that have elapsed between the arrival of the $i$-th \gls{SDU} in the $u$-th queue, and its successful delivery at the corresponding \gls{UE}.%
\footnote{For example, in Figure~\ref{fig:SDU_segmentation}, supposing that the \gls{SDU} \#5 is contained in \gls{TB} \#2 (colored in red), then its delay $D_u^{(5)}$ is the difference between the moment (i.e., \gls{TTI} index) at which \gls{TB} \#2 is received at the UE, and the moment at which \gls{SDU} \#5 has entered the queue (which in this example is $t_1$).}
If a total of $S_u$ \glspl{SDU} are transmitted to the $u^{th}$ UE in the run of an experiment, then the average \gls{SDU} latency experienced by the $u^{th}$ \gls{UE} is defined as
\begin{equation}
    \langle D_u \rangle
    = \frac{1}{S_u} \sum_{i=1}^{S_u} D_u^{(i)}.
\end{equation}
For practical reasons, we focus on constraining the delay of a statistically significant fraction of all \glspl{SDU}.
For instance, we may require that the fraction
\begin{equation}   \label{def:sigma_u}
	\sigma_u = \frac{\bigl| \bigl\{ i \colon D_u^{(i)} \leq \Delta \bigr\} \bigr|}{S_u}
\end{equation}
of \glspl{SDU} that satisfy some maximum target latency of $\Delta$ be larger than some threshold $\beta$ (i.e., $\sigma_u \geq \beta$).
We denote this quantity as \textbf{user satisfaction}.
For these activity and latency metrics to be realistic and free of exogenous artifacts, the packet arrival rate needs to be low enough so as to ensure that all queues remain stable, in the sense that the queue length is a stationary process. In simulations, we have tested this stability empirically, by verifying that queue lengths did not grow unbounded (leading to buffer saturation) for long runs of experiments.

\subsection{Modeling the \gls{PHY} layer}   \label{ssec:phy_modeling}

Each \gls{TB} that is conveyed from the \gls{BTS} to an active \gls{UE} at time $t$ has a certain probability $\epsilon(t)$ of being received incorrectly, which fluctuates over time depending on the channel conditions and channel estimation quality.

To define the stochastic process $\epsilon(t)$ we first adopt a simple first-order autoregressive fading model for the downlink channel gain $h(t)$, i.e.,
\begin{equation}
   h(t) = \rho \, h(t-1) + \sqrt{1-\rho^2} \, w(t)
\end{equation}
with parameter $\rho \in (0,1)$, where the white noise $w(t) \sim \mathcal{CN}(0,1)$ follows an independent complex normal distribution.
Inspired by the Jakes-Clarke one-ring model, the fading cross-correlation $\rho$ between two consecutive \glspl{TTI} can be expressed as (cf.~\cite[Eq.~(4)]{KuhKle08})\footnote{Note that this is not an exact application of the Jakes--Clarke model, since the latter describes a fading autocorrelation function described by $\mathsf{E}[ h(t) h(t-\Delta t)^*] = J_0(2\pi f_\mathrm{D} \Delta t)$. We limit ourselves to applying this expression of the autocorrelation function to a single \gls{TTI} slot.}
\begin{equation}
	\rho
	= J_0(2 \pi f_\mathrm{D} T)
\label{eq:rho_channel}
\end{equation}
where $J_0(x) = \frac{1}{2\pi} \int_{-\pi}^\pi e^{\jmath x \sin(t)} \mathrm{d}t$ denotes the Bessel function of the first kind of order zero, $T$ stands for the \gls{TTI} duration (one millisecond), $f_\mathrm{D} = f_\mathrm{c} v/c$ stands for the maximum Doppler spread, $f_\mathrm{c}$ for the carrier frequency, $v$ for the \gls{UE}'s velocity and $c$ for the speed of light.

\begin{figure}[ht!]
	\centering
	\includestandalone[scale=0.6]{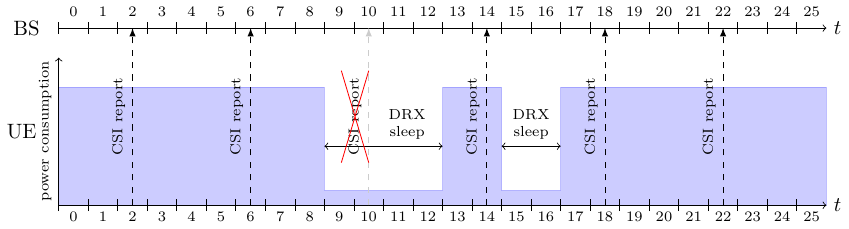}
	\caption{Pattern of periodic CSI reporting with interruptions due to the \gls{DRX} inactive mode.}
	\label{fig:CSI_reporting_pattern}
\end{figure}

The \glspl{UE} are assumed to be cognizant of the exact channel coefficient $h(t)$ at any time $t$, but report the current value of $h(t)$ only \emph{periodically} back to the \gls{BTS} over the \gls{PUCCH}.
Whenever a \gls{UE} is in the \gls{DRX} inactive state in a time slot wherein a \gls{CSI} report is due, said report is skipped, as illustrated in Figure~\ref{fig:CSI_reporting_pattern}. A similar approach can be found in~\cite{KuhKle08}.

\section{FORMULATION AS A PROTOCOL LEARNING PROBLEM}   \label{sec:DQN_architecture}
The interaction between the \gls{BTS} and \glspl{UE} through \glspl{CE} appended to \glspl{TB} forms a time-discrete, sequential decision-making process.
This process, governed by a \emph{policy}, is optimized using \gls{RL} at the \gls{BTS}.
Specifically, at each \gls{TTI}, the agent decides which \gls{MAC} \gls{CE} (if any) to send to each \gls{UE}, making independent decisions for each \gls{UE} (e.g., for 10 connected \glspl{UE}, 10 independent decisions are made).
Given the small action space, we use a value-based \gls{RL} algorithm, namely \gls{DQN}~\cite{Mnih13}.
While \gls{UE}-specific models are possible, a single agent model is more practical if it generalizes well across different \glspl{UE}.
To ensure that the agent can generalize well across \glspl{UE} with different traffic patterns and channel conditions, the state space includes \gls{UE}-specific features (e.g., queue length, age of packets in the queue) and cell-specific features (e.g., number of active \glspl{UE}, total queue length).
Thus, we trained a single \gls{DQN} agent with data from all \glspl{UE}.
This strengthens the agent's generalization ability, allowing it to make DRX signaling decisions based on each \gls{UE}'s local context.
The \gls{DQN} is updated based on the observed \gls{UE} traffic, enabling it to adapt to different traffic types and channel conditions for each \gls{UE}.
In the following, we define the Markovian state and action spaces available to the \gls{RL} agent at the \gls{BTS}.

\subsection{State vector definition}

We consider the following \gls{UE}-specific features to be available to the \gls{RL} agent as part of the state vector at time $t$:
\begin{itemize}
	\item	\textbf{Acknowledgment}: a binary variable indicating whether or not the downlink \gls{TB} transmitted at time $t-D_\mathrm{HARQ}$ was received correctly. We assume for simplicity that this acknowledgment arrives \emph{before} the \gls{BTS} policy execution at time $t$. The acknowledgment is therefore taken into account for computing the  \gls{BTS} action at time $t$ (see Section~\ref{ssec:actions}). $D_\mathrm{HARQ}$ denotes the \gls{SAW} \gls{HARQ} process delay, which for simplicity in our simulations we set to $D_\mathrm{HARQ}=1$.
	\item	\textbf{Scheduling indicator}: a binary variable indicating whether the \gls{UE} is scheduled for transmission at time $t$.
	\item	\textbf{Time until next \gls{DRX} long cycle}: a natural number that counts down the number of \glspl{TTI} remaining until the start of the next \gls{DRX} long cycle.
	\item	\textbf{Age of packets in queue}: the number of \glspl{TTI} that have elapsed since the oldest \gls{RLC} \gls{PDU} entered the downlink queue.
	\item	\textbf{Queue length}: the traffic volume (in number of bits) of the downlink queue for \gls{UE} $u$.
	\item	\textbf{Remaining time in current \gls{DRX} state}: the number of remaining \glspl{TTI} until the timers-based policy would trigger a \gls{DRX} state change.
	\item	\textbf{Resource usage}: a categorical variable with possible values in $\{0,1,2,3\}$ indicating how the radio resources were used in the previous step:
	\begin{itemize}
		\item	$0$: the \gls{UE} $u$ was scheduled and successfully transmitted a \gls{TB} containing at least one \gls{MAC} sub\gls{PDU} with payload.
		\item	$1$: the \gls{UE} $u$ was scheduled and successfully transmitted a \gls{TB} containing one or more \gls{MAC} sub\glspl{PDU}. None of them contained payload.
		\item	$2$: the \gls{UE} $u$ was scheduled, but the \gls{TB} transmission failed.
		\item	$3$: the \gls{UE} $u$ was not scheduled.
	\end{itemize}
\end{itemize}
In addition to these \gls{UE}-specific features, the following cell-specific features are also available to the \gls{RL} agent:
\begin{itemize}
	\item	\textbf{Number of active \glspl{UE}}: this indicates how many \glspl{UE} are in \gls{DRX} active state in the current time step.
	\item	\textbf{Total queue length}: the total number of bits at the queues of all active \glspl{UE} combined.
\end{itemize}
We collectively denote a vector containing all these features as $S_u(t)$. As can be seen from the above listing, there are eight numerical features and one quaternary categorical feature (resource usage). The latter being one-hot encoded in the neural implementation of the \gls{RL} policy, this results in $S_u(t)$ being represented by 12 inputs.

\subsection{Reward function}   \label{ssec:reward}

The reward function is designed to minimize \gls{UE} activity (modeled through the signal $W_u(t)$) while maintaining a minimum latency satisfaction $\sigma_u(t)$.
In our experiments, we strive to achieve that a target fraction $\beta$ of \glspl{SDU} meets a given target maximum latency $\Delta$. For this purpose, at any time $t$ we look back at a fixed number $N = 20$ of the most recently transmitted \glspl{SDU} and compute the metric $\sigma_u$ based on~\eqref{def:sigma_u} over that sample. 
We then define the reward function $R_u(t)$ of the $u$-th \gls{UE} as the ``satisfaction gap'' $\sigma_u(t) - \beta$ whenever $\sigma_u(t)<\beta$.
Otherwise, the reward function is equal to the sleep state indicator $1 - W_u(t)$. That is,
\begin{equation*}
	R_u(t)
	= \begin{cases}
		\sigma_u(t) - \beta & \text{if $\sigma_u(t) < \beta$} \\
		1 - W_u(t)          & \text{if $\sigma_u(t) \geq \beta$}
	\end{cases}
\end{equation*}
We further define the average \emph{cumulative reward} per \gls{UE} as the sum of reward values accumulated over the course of an episode, averaged over the set of active \glspl{UE}, of which there is a fixed number $U \in \{1, \dotsc, 9\}$ in any given episode, i.e., $\frac{1}{U} \sum_{u=1}^{U} \sum_{t=1}^{N_\mathrm{t}} R_u(t)$.

\subsection{Actions}   \label{ssec:actions}

The definition of the action space for the RL agent depends on whether we are training in a 5G NR-compliant scenario or not.
As detailed in Section~\ref{sec:simulation_results} below, we consider two cases, with respective action space sizes $|\mathcal{A}| = 2$ (compatible with pre Rel-17 5G NR) and $|\mathcal{A}| = 7$ (compatible with Rel-17 5G NR and beyond):
\begin{itemize}
	\item	Case 1 ($|\mathcal{A}| = 2$):
	\begin{itemize}
		\item	If $A_u(t) = 1$: the \gls{BTS} includes a \texttt{Long\_DRX\_command\_MAC\_CE} (\gls{LCID} 59) in the first MAC SDU of the next \gls{TB}. This commands the \gls{UE} to start a long DRX inactivity period immediately.
		\item	If $A_u(t) = 0$: no \gls{MAC} \gls{CE} is included. Hence, the \gls{TB} only contains a header and data payload.
	\end{itemize}
	\item	Case 2 ($|\mathcal{A}| = 7$): same as Case~1, except that the inactivity duration can be selected with more granularity via a custom \gls{CE}. The value of $A_u(t) \in \{0, 2, 4, 6, 8, 10, 12\}$ denotes the duration of this inactivity as a number of \glspl{TTI}. This is similar to the \emph{PDCCHSkippingDurationList} recently introduced by 3GPP in Rel-17.
\end{itemize}
Note that, for simplicity of Case 1, we do not consider the usage of the Short DRX command, which is rarely used in practical deployments.

\subsection{\gls{ERM} management}

The transition tuples stored into the \gls{ERM} are defined as $(S_u(t),A_u(t),R_u(t),S_u(t+1))$, where $u \in \{1,\dotsc,U\}$.
Note that, in contrast to the textbook setting of single-agent \gls{RL}, in our setting the \gls{BTS} policy is not executed at \emph{every} \gls{TTI} index $t$, but only \emph{conditionally} when the receiving \gls{UE} is known to be active (listening).
Otherwise, the \gls{UE} is known to be inactive and no policy is executed.
In the latter case (when the \gls{UE} is inactive), we deviate from the classic \gls{DQN} implementation as follows:
\begin{itemize}
	\item We force the \gls{BTS} policy to choose a null action (by hard-coding a switch).
	\item We do not store the corresponding transition into the \gls{ERM}. This is for practical reasons due to the limited capacity of the \gls{ERM}. Storing transitions containing the null action might overwrite other \emph{more interesting} RL-guided transitions.
\end{itemize}

\section{EXPERIMENTAL VALIDATION}   \label{sec:simulation_results}
\begin{table}[ht]
\caption{System settings}
\label{tab:sim_config}
\begin{center}
\begin{tabular}{|l|l|}
\hline
TTI & 1 ms
\\ \hline
Bandwidth & 100 MHz
\\ \hline
Effective BW for DL payload & 72 MHz
\\ \hline
Fading cross-correlation $\rho$ & 0.99
\\ \hline
CSI update period $T_\mathrm{CSI}$ & 10 ms
\\ \hline
Traffic type & XR (see Table \ref{table:XR_traffic})
\\ \hline
Number of \glspl{UE} & 1 to 9
\\ \hline
SNR & 10 dB
\\ \hline
\textit{drxInactivityTimer} & 8 ms
\\ \hline
\textit{drxOnDurationTimer} & 8 ms
\\ \hline
\textit{drxLongCycle} & 16 ms
\\ \hline
\end{tabular}
\end{center}
\end{table}

We assess the performance of the proposed \gls{RL}-based \gls{MAC} signaling \gls{DRX} scheme in a system-level simulation where one \gls{BTS} serves several \glspl{UE}.
Table~\ref{tab:sim_config} collects all configuration details, where the value of the \emph{drxLongCycle} has been aligned with the packet interarrival time for \gls{XR} traffic (other traffic types, such as non-GBR data might use values closer to 160\,ms, voice might use 40\,ms, etc.).
Finally, we compare our method against the following baselines:
\begin{enumerate}
	\item	\textbf{Always ON}: \glspl{UE} listen decode \gls{PDCCH} on all \glspl{TTI}.
	\item	\textbf{Timers-based}: the \glspl{UE} \gls{DRX} activity and inactivity periods are fully controlled via \gls{RRC} timers as per 3GPP standards.
	\item	\textbf{Na\"ive policy}: in addition to \gls{RRC} timers, the \gls{BTS} also instructs \glspl{UE} via legacy \glspl{CE} (LCIDs) to commence inactive periods as soon as their traffic queue empties.
	\item	\textbf{Random policy}: in addition to the \gls{RRC} timers, the \gls{BTS} sends legacy \glspl{CE} with probability $1/2$ each \gls{TTI}.
\end{enumerate}
These baselines were benchmarked against the following \gls{RL}-based solutions:
\begin{enumerate}
	\item	\textbf{\gls{RL} policy ($|\mathcal{A}|=2$)}: as in the timers-based baseline (item 2 above), but the BTS now decides whether or not to send legacy \gls{5G NR}-compliant \gls{MAC} \glspl{CE} based on a \gls{RL} policy.
	\item	\textbf{\gls{RL} policy ($|\mathcal{A}|=7$)}: as in the previous \gls{RL} scheme, but the BTS now decides to send custom \glspl{CE} that instruct the \gls{UE} to commence inactive periods of variable duration based on a \gls{RL} policy (see Subsection~\ref{ssec:actions} for details).
\end{enumerate}
All \glspl{UE} share the same XR traffic model and RRC timer configurations, with the DRX inactivity timer set to 8 ms following commercial products, and the DRX long cycle duration set to 16 ms for alignment with the XR packet interarrival time.
The na\"ive, random and \gls{RL} policies are always \emph{stabilized} by a hard-coded rule that prohibits the transmission of any \gls{CE} in case of queue saturation. This is a trivial fix to improve the performance of these policies, and it felt only fair to include it here.

\subsection{Hyper-parameter tuning}   \label{sec:hyperparameter_tuning}

\begin{table}
\caption{Reinforcement Learning hyperparameters}
\label{tab:rl_hp}
\begin{center}
\setlength\tabcolsep{1.5mm}
\begin{tabular}{|l|l|}
\hline
\gls{ERM} size $M$ (number of transition tuples) & $10^5$
\\ \hline
Batch size $B$ (number of transition tuples) & 256
\\ \hline
Size $|\mathcal{A}|$ of action space & $2$ or $7$
\\ \hline
Number of hidden layers & 1
\\ \hline
Neurons per hidden layer & $40$
\\ \hline
Activation function (hidden layers) & rect.~linear unit
\\ \hline
Activation function (output layer) & softmax
\\ \hline
Input layer size (for a history size of 3 \glspl{TTI}) & $12 \times 3 = 36$
\\ \hline
Output layer size (equals action space size) & 2 or 7
\\ \hline
Loss function & Huber loss (param.~1)
\\ \hline
Number of inferences between model updates & 100
\\ \hline
Number $N$ of SDUs for computing metric $\sigma_u$ & 20
\\ \hline
Discount factor $\gamma$ & $1$
\\ \hline
Learning rate $\mathrm{LR}$ & \makecell[l]{$10^{-2} \rightarrow 10^{-5}$ (tuning) \\ $10^{-3}$ (training)}
\\ \hline
Number of indep.\ training sessions $N_\mathrm{runs}$ & 8 (tuning), 30 (training)
\\ \hline
Number $N_\mathrm{ep}$ of episodes per run & \makecell[l]{750 (training) \\ 250 (evaluation)}
\\ \hline
Number $N_\mathrm{t}$ of time steps per episode & 8000
\\ \hline
Epsilon-greedy decay & $0.8 \rightarrow 10^{-6}$
\\ \hline
Delay stringency $\beta$ & $0.95$
\\ \hline
\end{tabular}
\end{center}
\end{table}

Most of the hyper-parameter values we have selected are commonly used across the \gls{RL} literature.
However, we have opted for a slightly larger batch size $B$ of 256 to reduce noise during gradient descent updates.
In addition, we have also fine-tuned the learning rate and the discount factor by evaluating the performance of the proposed scheme across $N_\mathrm{runs}=30$ randomized training runs for each candidate hyper-parameter value.
Then, we select the one yielding the highest average (over the training runs) cumulative reward. Each training run consists of $N_\mathrm{ep}=750$ independent episodes with $N_t=8000$ time steps each, where the first episodes are largely devoted to exploration. Specifically, the exploration--exploitation balance is controlled by a parameter $\epsilon \in [0,1]$ which is exponentially decreased from $0.8$ to $10^{-6}$, in steps every 30 episodes, reaching its final value of $10^{-6}$ after 300 episodes, where it stops decaying.

Once the best configuration is selected, we carry out extensive training ($N_\mathrm{runs}=30$) to derive the \gls{RL} model, selecting the one with the highest historic cumulative reward.

\subsection{Simulation results}

\begin{figure}[!ht]
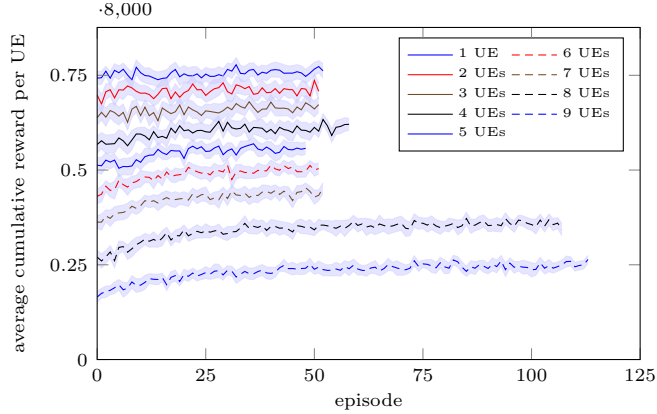

\centering
\includestandalone{pgfplots/CR_training_convergence_actions2}
\caption{Convergence of the cumulative reward for $|\mathcal{A}|=2$ and a variable number of \glspl{UE}, averaged over $N_\mathrm{runs}=30$ independent runs of the training. Traffic statistics, system model and parameters are as described in Tables~\ref{table:XR_traffic}, \ref{tab:sim_config}, \ref{tab:rl_hp}.}
\label{fig:CR_training_convergence_actions2}
\end{figure}

\begin{figure}[!ht]
\centering
\includestandalone{pgfplots/US_training_convergence_actions2}
\caption{Convergence of the user satisfaction corresponding to the same training runs as in Figure~\ref{fig:CR_training_convergence_actions2}.}
\label{fig:US_training_convergence_actions2}
\end{figure}

\begin{figure}[!ht]
\centering
\includestandalone{pgfplots/CR_training_convergence_actions7}
\caption{Convergence of the cumulative reward for $|\mathcal{A}|=7$ and a variable number of \glspl{UE}, averaged over $N_\mathrm{runs}=30$ independent runs of the training. Traffic statistics, system model and parameters are as described in Tables~\ref{table:XR_traffic}, \ref{tab:sim_config}, \ref{tab:rl_hp}.}
\label{fig:CR_training_convergence_actions7}
\end{figure}

\begin{figure}[!ht]
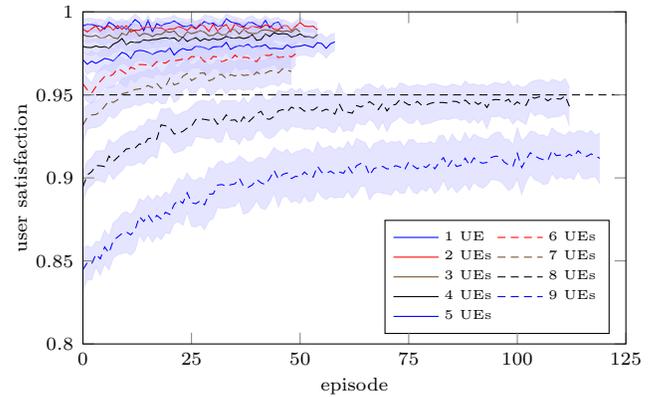

\centering
\includestandalone{pgfplots/US_training_convergence_actions7}
\caption{Convergence of the user satisfaction corresponding to the same training runs as in Figure~\ref{fig:CR_training_convergence_actions7}.}
\label{fig:US_training_convergence_actions7}
\end{figure}

In Figures~\ref{fig:CR_training_convergence_actions2} and \ref{fig:CR_training_convergence_actions7}, we show the convergence of the average cumulative reward per \gls{UE}, as previously defined in Section~\ref{ssec:reward}, for the cases $|\mathcal{A}| = 2$ and $|\mathcal{A}| = 7$, respectively. Note that the $y$ axes indicate the cumulative reward per \gls{UE} and per \gls{TTI}, and the indication ``$\cdot 8,000$'' at the axis top is a reminder that $y$-axis values need to be multiplied by the corresponding factor to obtain the cumulative reward per episode (since there are 8000 \glspl{TTI} per episode, cf.~Table~\ref{tab:rl_hp}). Each curve represents a mean value over $N_\mathrm{runs} = 30$ independent training sessions, plotted along with a 95\% $t$-student confidence interval.
We intended to learn policies that are robust to variations in the number of \glspl{UE}.
To achieve this, we randomly varied (from $1$ to $9$) the number of active \glspl{UE} between episodes in an i.i.d.\ fashion (with more weight given to scenarios with $8$ and $9$ \glspl{UE}).
The training curves in Figure~\ref{fig:CR_training_convergence_actions2} have been arranged separately based on the number of active \glspl{UE}, offering a more distinct portrayal of trends.
The randomness in the activation of \glspl{UE} explains the different curve lengths visible in Figures \ref{fig:CR_training_convergence_actions2} through \ref{fig:US_training_convergence_actions7}.
Figures~\ref{fig:US_training_convergence_actions2} and \ref{fig:US_training_convergence_actions7} show the evolution of user satisfaction for the same training experiments as in Figures~\ref{fig:CR_training_convergence_actions2} and \ref{fig:CR_training_convergence_actions7}, respectively. The dashed line indicates the target delay threshold $\beta = 0.95$ that the user satisfaction should ideally exceed. We see that with the \gls{5G NR} compliant scheme ($|\mathcal{A}| = 2$) we can only serve up to 3 \glspl{UE} with satisfactory delays, whereas the more granular signaling ($|\mathcal{A}| = 7$) allows to serve up to 8 \glspl{UE}.

\begin{figure}[!ht]
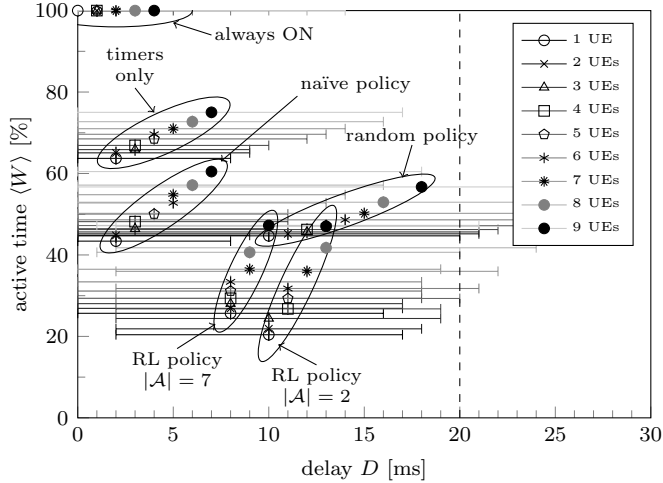

\centering
\includestandalone{pgfplots/Eval_rho_0.99_Delay_vs_OnTime_1_2_4_6}
\caption{Tradeoff between average active time and delay quantiles for a varying number of \glspl{UE} and different \gls{DRX} schemes in comparison, with same parameters as in Figure~\ref{fig:CR_training_convergence_actions2}. For every configuration, the average delay is shown at the marker position (see legend) together with a horizontal bar with right and left whiskers, which are placed at the 5-th and 95-th delay percentile, respectively.}
\label{fig:uptime-delay_tradeoff}
\end{figure}

In Figure~\ref{fig:uptime-delay_tradeoff}, we illustrate the average performance of the different baselines and trained policies in the activity--delay trade-off plane.
Here, the $x$-axis denotes the \gls{SDU} delay $D = \tfrac{1}{U} \sum_{u=1}^U D_u$ (in ms), and the boxplots depict the median, and 5-th/95-th percentiles of the delay distribution for each solution.
Similarly, the $y$-axis denotes the fraction of active time $\langle W \rangle = \tfrac{1}{U} \sum_{u=1}^U \langle W_u \rangle$ (as a percentage), where $U \in \{1,\dotsc,9\}$ is the number of \glspl{UE}.
In this plot, we compare the performance of RL-based DRX for $|\mathcal{A}|=2$ and $|\mathcal{A}|=7$ against several baselines.

For each configuration, a horizontal boxplot shows the median (corresponding to the plot mark, as in the legend) as well as left and right whiskers that correspond to the 5-th and 95-th percentile of the empirical \gls{SDU} delay distribution, respectively.
The maximum admissible delay (20 ms) is displayed by a dashed vertical delimiting line.
Ellipses are a visual help to distinguish clusters of boxplots that correspond to different configurations.

Note that the simulation results for the \gls{RL} agents ($|\mathcal{A}|=2$ and $|\mathcal{A}|=7$) correspond to a single agent, respectively, which is exposed during training to an environment with a number of active \glspl{UE} that varies randomly between 1 and 9 (as in the context of Figure~\ref{fig:CR_training_convergence_actions2}), and which is then evaluated for a \emph{fixed} number of active \glspl{UE} (this value ranging from 1 to 9). The other schemes correspond to an Always ON operation (no \gls{DRX}), timers only, as well as two baseline schemes: the na\"ive policy consists in having every \gls{UE} send a \gls{CE} as soon as its queue is empty. The random policy decides whether or not to send a \gls{CE} based on a Bernoulli-0.5 variable (fair coin flip).
Recall that the na\"ive, random and \gls{RL} policies are always \emph{stabilized} to prevent queue saturation, as explained at the beginning of Section~\ref{sec:simulation_results}.
 
We see that the \gls{RL}-based \gls{CE}-controlled \gls{DRX} outperforms all baselines in terms of activity reduction (and thus also the \gls{DRX}-related power savings) by a substantial margin, while meeting the 20\,ms latency target for most configurations: by keeping track of the locations of right whiskers on the boxplots (95-th percentiles of \gls{SDU} delay) in Figure~\ref{fig:uptime-delay_tradeoff}, we see that up to 5 \glspl{UE} can be served with satisfactory delays with a \gls{5G NR} compatible \gls{RL} policy ($|\mathcal{A}|=2$) whereas 8 \glspl{UE} can be served with the more granular policy ($|\mathcal{A}|=7$).

\begin{figure}[!ht]
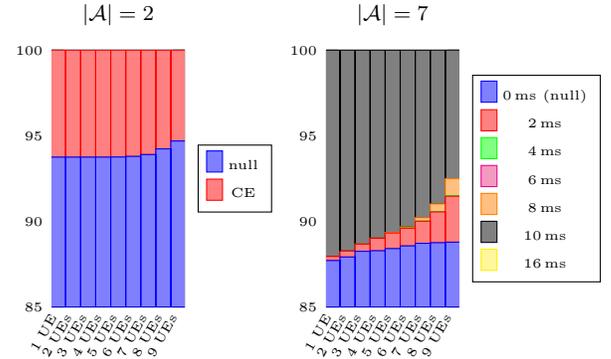

	\centering
	\includestandalone{pgfplots/action_frequencies_5G}
	\includestandalone{pgfplots/action_frequencies_6G}
	\caption{Relative frequencies (in percentages) of the selected actions.}
	\label{fig:action_frequencies}
\end{figure}
Finally, Figure~\ref{fig:action_frequencies} shows the relative frequencies of selected actions. We observe that the null action (corresponding to 0\,ms) is by far the most frequent action. We can also appreciate how a richer action space ($|\mathcal{A}|=7$ instead of $|\mathcal{A}|=2$) seems to be exploited by the \gls{RL} agent by privileging certain actions over others. In our case, it is mostly those actions corresponding to 10\,ms, 2\,ms and 8\,ms which are utilized with non-negligible probability. This is likely dependent on the specificities of the \gls{XR} traffic pattern and some of our parameter choices, e.g., the value of 16\,ms for the \textit{drxLongCycle}.\footnote{This relatively short cycle is chosen to align with the extremely low \gls{PDB} of \gls{XR} traffic.}

\section{CONCLUSION}
This paper has shown that a signaling approach to DRX can augment a \gls{5G NR}-compliant and timers-based conventional \gls{DRX} scheme.
We have also shown that \gls{RL} techniques can discover deterministic DRX control policies that leverage \gls{MAC} \glspl{CE} to trigger additional inactivity periods with judicious timing.
Simulation results have shown the potential of both \glspl{CE} and \gls{RL} to achieve substantial energy savings beyond timers-based \gls{DRX} baselines while still fulfilling strict QoS latency requirements.
This opens up the possibility of fine-tuning the operating regime of UE-side PDCCH detection in the activity--latency trade-off plane with a very fine granularity.
Driven by economics and sustainability goals, the importance of DRX and other energy-saving features will only grow on the path toward 6G, and fine-tuning its operating point can maximize such savings.
In addition, \gls{ML}-aided policies, while remaining standard compliant, also have the potential to optimize \gls{DRX} by leveraging other contextual data, such as scheduling decisions, traffic statistics, and obscure cross-UE interdependencies.
Designing control policies capable of digesting such vast amounts of data in real-time is challenging.
The simulation results presented in this paper suggest that automated policy discovery methods like the one described here are a fast and cost-effective way to achieve energy savings in 5G and beyond.


\bibliographystyle{IEEEtran}
\bibliography{IEEEabrv,refs}

\begin{thebibliography}{10}
\providecommand{\url}[1]{#1}
\csname url@samestyle\endcsname
\providecommand{\newblock}{\relax}
\providecommand{\bibinfo}[2]{#2}
\providecommand{\BIBentrySTDinterwordspacing}{\spaceskip=0pt\relax}
\providecommand{\BIBentryALTinterwordstretchfactor}{4}
\providecommand{\BIBentryALTinterwordspacing}{\spaceskip=\fontdimen2\font plus
\BIBentryALTinterwordstretchfactor\fontdimen3\font minus
  \fontdimen4\font\relax}
\providecommand{\BIBforeignlanguage}[2]{{%
\expandafter\ifx\csname l@#1\endcsname\relax
\typeout{** WARNING: IEEEtran.bst: No hyphenation pattern has been}%
\typeout{** loaded for the language `#1'. Using the pattern for}%
\typeout{** the default language instead.}%
\else
\language=\csname l@#1\endcsname
\fi
#2}}
\providecommand{\BIBdecl}{\relax}
\BIBdecl

\bibitem{Valcarce21}
A.~Valcarce and J.~Hoydis, ``Toward joint learning of optimal {MAC} signaling
  and wireless channel access,'' \emph{IEEE Transactions on Cognitive
  Communications and Networking}, vol.~7, no.~4, pp. 1233--1243, 2021.

\bibitem{Dahlman24}
E.~Dahlman, S.~Parkvall, and J.~Sköld, \emph{5G/5G-Advanced. The new
  generation wireless access technology (3rd ed.)}.\hskip 1em plus 0.5em minus
  0.4em\relax Elsevier Ltd, 2024.

\bibitem{Ruy2020}
Y.-N.~R. Li, M.~Chen, J.~Xu, L.~Tian, and K.~Huang, ``Power saving techniques
  for {5G} and beyond,'' \emph{IEEE Access}, vol.~8, pp. 108\,675--108\,690,
  2020.

\bibitem{zte18}
\BIBentryALTinterwordspacing
Z.~Corporation, ``Consideration on {UE} power consumption model and preliminary
  evaluation results,'' Nov 2018. [Online]. Available:
  \url{https://portal.3gpp.org/ngppapp/TdocList.aspx?meetingId=18807}
\BIBentrySTDinterwordspacing

\bibitem{5GMAC}
3GPP, ``{TS} 38.321 {NR} {M}edium {A}ccess {C}ontrol ({MAC}) protocol
  specification,'' {3rd Generation Partnership Project (3GPP)}, Technical
  Specification (TS), {TSG-RAN1\#48 R1-070674}.

\bibitem{9771950}
A.~A. Esswie, ``Power saving techniques in 3gpp 5g new radio: A comprehensive
  latency and reliability analysis,'' in \emph{2022 IEEE Wireless
  Communications and Networking Conference (WCNC)}, 2022, pp. 66--71.

\bibitem{Muda19}
M.~Memon, M.~Maheshwari, D.~Shin, A.~Roy, and N.~Saxena, ``Deep‐{DRX}: A
  framework for deep learning–based discontinuous reception in {5G} wireless
  networks,'' \emph{Transactions on Emerging Telecommunications Technologies},
  vol.~30, 03 2019.

\bibitem{Zhou2019}
J.~Zhou, G.~Feng, T.~S.~P. Yum, M.~Yan, and S.~Qin, ``{Online learning-based
  discontinuous reception (DRX) for machine-type communications},'' \emph{IEEE
  Internet of Things Journal}, vol.~6, no.~3, pp. 5550--5561, 2019.

\bibitem{Ericsson21}
P.~Bruhn and G.~Bassi, ``Machine learning based {C-DRX} configuration
  optimization for {5G},'' in \emph{Mobile Communication - Technologies and
  Applications; 25th ITG-Symposium}, 2021, pp. 1--6.

\bibitem{Akyi22}
I.~Akyildiz and H.~Guo, ``Wireless communication research challenges for
  {E}xtended {R}eality {(XR)},'' \emph{ITU Journal on Future and Evolving
  Technologies (ITU J-FET)}, vol.~3, 04 2022.

\bibitem{PetGapParMarPed22}
\BIBentryALTinterwordspacing
V.~Petrov, M.~Gapeyenko, S.~Paris, A.~Marcano, and K.~I. Pedersen,
  ``Standardization of {E}xtended {R}eality {(XR)} over {5G} and {5G-Advanced}
  {3GPP} {N}ew {R}adio,'' 2022. [Online]. Available:
  \url{https://arxiv.org/abs/2203.02242}
\BIBentrySTDinterwordspacing

\bibitem{KuhKle08}
A.~Kuhne and A.~Klein, ``Throughput analysis of multi-user {OFDMA}-systems
  using imperfect {CQI} feedback and diversity techniques,'' \emph{IEEE Journal
  on Selected Areas in Communications}, vol.~26, no.~8, pp. 1440--1450, 2008.

\bibitem{Mnih13}
\BIBentryALTinterwordspacing
V.~Mnih, K.~Kavukcuoglu, D.~Silver, A.~Graves, I.~Antonoglou, D.~Wierstra, and
  M.~A. Riedmiller, ``Playing {A}tari with deep reinforcement learning,''
  \emph{CoRR}, vol. abs/1312.5602, 2013. [Online]. Available:
  \url{http://arxiv.org/abs/1312.5602}
\BIBentrySTDinterwordspacing

\end{thebibliography}

\newpage
 
\begin{IEEEbiographynophoto}
{Adriano Pastore,}~(Senior Member, IEEE)~is a Senior Researcher at the Centre Tecnològic de Telecomunicacions de Catalunya, within the Research Unit on Information and Signal Processing for Intelligent Communications. He received a Diplôme de l’École Centrale Paris (now CentraleSupélec) in 2006 and a Dipl.-Ing.\ degree in electrical engineering in 2009 from the Technical University of Munich, and obtained his PhD from the Universitat Politècnica de Catalunya in 2014. From 2014 to 2016 he has been a postdoctoral researcher at École Polytechnique Fédérale de Lausanne (EPFL) with Prof.~Michael Gastpar.

His topics of interest lie mainly in the fields of information theory and signal processing for wireless communications, machine learning for communications, physical-layer network coding, protocol learning, quantum key distribution, and privacy--utility tradeoffs.
\end{IEEEbiographynophoto}

\begin{IEEEbiographynophoto}
{Adrián Agustín,}~(Senior Member, IEEE)~is a Senior Researcher at the Centre Tecnològic de Telecomunicacions de Catalunya (CTTC), within the Research Unit on Information and Signal Processing for Intelligent Communications (ISPIC).He received the M.S. and Ph.D in Telecommunication from Universitat Politècnica de Catalunya (UPC), Barcelona, Spain in 2000 and 2008, respectively. From 2008 to 2019 he was a research associate at the SPCOM group at UPC working in the different research areas of wireless communications. In April, 2021 he joined the CTTC.

He is interested in different research areas for developing next generation wireless communications at L1, L2, in particular with : Extremely Large Antenna Arrays (ELAA), Deep-Reinforcement Learning (DRL) and Random Access Networks.
\end{IEEEbiographynophoto}

\begin{IEEEbiographynophoto}
{Álvaro Valcarce,}~(Senior Member, IEEE)~is Head of Department on Wireless AI/ML at Nokia Bell Labs, France. His research is focused on the application of machine learning techniques to L2 and L3 wireless problems for the development of technologies beyond 5G. He is especially interested on the potential of multiagent reinforcement learning for emerging novel L2 signaling protocols, as well as on the usage of Bayesian optimization for RRM problems. His background is on cellular networks, computational electromagnetics, optimization algorithms, and machine learning.
\end{IEEEbiographynophoto}

\vfill\pagebreak

\end{document}